\documentstyle[sprocl,epsfig]{article}

\def\LQCD{$\Lambda_{\mbox{\scriptsize QCD}}$ }
\def\as{{\alpha}_{\mbox{\scriptsize s}}}
\def\asl{\alpha^{(\ell)}_{\mbox{\scriptsize s}}}
\def\aslt{\widetilde{\alpha}^{(\ell)}_{\mbox{\scriptsize s}}}
\def\aani{\alpha^{\mbox{\scriptsize (1)}}_{\mbox{\scriptsize an}}}
\def\aanl{\alpha^{(\ell)}_{\mbox{\scriptsize an}}}

\def\aanlt{\widetilde{\alpha}^{(\ell)}_{\mbox{\scriptsize an}}}
\def\aanit{\widetilde{\alpha}^{\mbox{\scriptsize (1)}}
    _{\mbox{\scriptsize an}}}
\def\myaani{{^{\mbox{\tiny N}}}\!\aani}

\def\myaanl{{^{\mbox{\tiny N}}}\!\aanl}

\def\myaanlt{{^{\mbox{\tiny N}}}\!\aanlt}
\def\myaanit{{^{\mbox{\tiny N}}}\!\aanit}
\newcommand{\sfa}{\mbox{\sf A}}

\begin{document}

\title{NEW MODEL FOR THE QCD \\ ANALYTIC RUNNING COUPLING}

\author{A.\ V.\ NESTERENKO}

\address{Department of Physics, Moscow State University,\\
Vorobiyovy Hills, Moscow, 119899, Russian Federation\\
E-mail: nesterav@thsun1.jinr.ru}

\maketitle\abstracts{As an elaboration of the analytic approach to
Quantum Chromodynamics (QCD), a new model for the QCD analytic
running coupling is proposed. A self-evident advantage of the new
analytic running coupling (NARC) is that it incorporates the infrared
enhancement and the ultraviolet asymptotic freedom in a single
expression. It is essential that additional parameters are not
introduced in the theory. The absence of unphysical singularities in
the physical region $q^2>0$ and a fairly well loop and scheme
stability are the remarkable features of the NARC. By making use of
practically the same values of $\Lambda_{\mbox{\tiny QCD}}$ in the
approach developed, one succeeded in description of various physical
phenomena, from quark confinement to the $\tau$ lepton decay. This
undoubtedly implies that the new analytic running coupling
substantially involves both the nonperturbative and the perturbative
behavior of Quantum Chromodynamics.}

\section{Introduction}

     The discovering of the asymptotic freedom phenomenon in Quantum
Chromodynamics (QCD) led to the wide using of the perturbation
theory.  The latter gives good results in the ultraviolet (UV)
region, but the perturbation theory is absolutely unapplicable in the
infrared (IR) region. So, for the description of a number of physical
phenomena one needs to use the nonperturbative methods. The current
consideration relies on the analytic approach to QCD.

     The analytic approach to Quantum Field Theory (QFT) is a
nonperturbative method which is based solely on the first principles
of the local QFT. First it was formulated on the example of Quantum
Electrodynamics in the late 1950's in the papers.\cite{Redm,BLS} The
basic idea of this approach is the explicit imposition of the
causality condition, which implies the requirement of the analyticity
in the $q^2$ variable for the relevant physical quantities.  This
approach has recently been extended to QCD \cite{ShSol} and applied
to the ``analytization'' of the perturbative series for the QCD
observables.\cite{SolSh} The term ``analytization'' means the
recovering of the proper analytic properties in the $q^2$ variable by
making use of the K\"all\'en--Lehmann spectral representation
\begin{equation}
\Bigl\{\sfa(q^2)\Bigr\}_{\mbox{$\!${\small an}}} \equiv
\int_{0}^{\infty}\! \frac{\varrho(\sigma)}{\sigma+q^2}\, d\sigma,
\end{equation}
where the spectral density is determined by the initial
(perturbative) expression for some quantity {\sf A}$(q^2)$:
\begin{equation}
\varrho(\sigma) \equiv \frac{1}{2 \pi i} \lim_{\varepsilon \to 0_{+}}
\Bigl[\sfa(-\sigma-i\varepsilon)-\sfa(-\sigma+i\varepsilon)\Bigr],
\quad \sigma \ge 0.
\end{equation}

\section{New Analytic Running Coupling}

     First of all, let us consider the renormalization group (RG)
equation for the invariant charge $g(\mu)$
\begin{equation}
\label{RGGen}
\frac{d\,\ln \bigl[g^2(\mu)\bigr]}{d\,\ln\mu^2} =
\beta\Bigl(g(\mu)\Bigr).
\end{equation}
In accordance with the perturbative approach one expands the
$\beta$ function on the right hand side of Eq.\ (\ref{RGGen}) as a
power series
\begin{equation}
\label{BetaPertSer}
\beta\Bigl(g(\mu)\Bigr) = - \left\{
\beta_{0}\left[\frac{g^2(\mu)}{16 \pi^2}\right] +
\beta_{1}\left[\frac{g^2(\mu)}{16 \pi^2}\right]^2 + \ldots \right\},
\end{equation}
where $\beta_{0} = 11 - 2\,n_{\mbox{\scriptsize f}}\,/\,3,\,
\beta_{1}=102 - 38 \,n_{\mbox{\scriptsize f}}\,/\,3$, and
$n_{\mbox{\scriptsize f}}$ is the number of active quarks.
Introducing the standard notations $\as(\mu^2)= g^2(\mu)/(4\pi)$ and
$\widetilde{\alpha}(\mu^2) = \alpha(\mu^2)\, \beta_{0}/(4\pi)$, one
can reduce the RG equation at the $\ell$-loop level to the form
\begin{equation}
\frac{d\,\ln\bigl[\aslt(\mu^2)\bigr]}{d\,\ln \mu^2} = -
\sum_{j=0}^{\ell-1} \beta_j
\left[\frac{\aslt(\mu^2)}{\beta_{0}}\right]^{j+1}.
\end{equation}
It is well-known that the solution to this equation has unphysical
singularities at any loop level. So, there is the ghost pole at the
one-loop level, and the account of the higher loop corrections just
introduces the additional singularities of the cut type into the
expression for the running coupling. But we know from the first
principles that the QCD running coupling must have the correct
analytic properties in the $q^2$ variable, namely, there should be
the only left cut-off along the negative semiaxis of $q^2$.

     So, what is missing in the $\beta$ function perturbative
expansion?  How one could try to improve the situation? This
objective can be achieved by involving into consideration the
analytic properties in the $q^2$ variable of the RG equation. The
perturbative expansion of the $\beta$ function as a power series
(\ref{BetaPertSer}) leads to the violation of the correct analytic
properties of the RG equation (\ref{RGGen}). Thus, one can improve
the $\beta$ function perturbative expansion by applying the
analytization procedure to it.\cite{PRD} This results in the
equation
\begin{equation}
\label{AnRGEq}
\frac{d\,\ln\bigl[\myaanlt(\mu^2)\bigr]}{d\,\ln \mu^2} = - \left\{
\sum_{j=0}^{\ell-1} \frac{\beta_{j}}{\beta_{0}^{j+1}}
\Bigl[\aslt(\mu^2)\Bigr]^{j+1}\right\}_{\mbox{$\!\!$\small an}},
\end{equation}
which solution is by definition \cite{PRD,NARCSTR} the $\ell$-loop
new analytic running coupling, $\myaanl(q^2)$. In fact, the solution
to Eq.\ (\ref{AnRGEq}) is determined up to a constant factor, but
this ambiguity can easily be avoided by imposing the physical
condition of the asymptotic freedom $\myaanl(q^2) \to \asl (q^2)$,
when $q^2 \to \infty$. At the one-loop level Eq.\ (\ref{AnRGEq}) can
be integrated explicitly with the result
\begin{equation}
\label{NARCOneLoop}
\myaani(q^2) = \frac{4\pi}{\beta_0}\,\frac{z-1}{z\,\ln z}, \qquad
z=\frac{q^2}{\Lambda^2}.
\end{equation}
\begin{figure}[ht]
\vspace*{-5mm}
\centerline{\epsfig{file=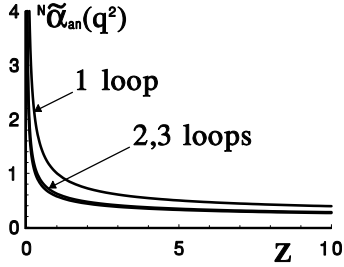, width=45mm}}
\caption{The new analytic running coupling at different loop levels,
$z=q^2/\Lambda^2$.}
\label{Fig:NARC}
\end{figure}

\noindent
At the higher loop levels there is only the integral representation
for the new analytic running coupling (NARC). So, at the $\ell$-loop
level we have
\begin{equation}
\label{NARCHighLoop}
\myaanlt(q^2) = \frac{C^{(\ell)}}{z}\, \exp\!\left[
\int_{0}^{\infty}\!{\cal R}^{(\ell)}(\sigma)\,
\ln\!\left(\frac{\sigma+z}{\sigma+1}\right)
\frac{d\sigma}{\sigma}\right],
\end{equation}
where
\begin{equation}
{\cal R}^{(\ell)}(\sigma) = \frac{1}{2 \pi i}
\lim_{\varepsilon \to 0_{+}}
\sum_{j=0}^{\ell-1} \frac{\beta_{j}}{\beta_{0}^{j+1}}
\biggl\{\Bigl[\aslt(-\sigma-i\varepsilon)\Bigr]^{j+1}
-\Bigl[\aslt(-\sigma+i\varepsilon)\Bigr]^{j+1}\biggr\},
\end{equation}
and the normalization coefficients are $C^{(1)}=1.0000,\,$
$C^{(2)}=0.5723,\,$ $C^{(3)}=0.6220$. The Figure \ref{Fig:NARC} shows
the NARC (\ref{NARCHighLoop}) at the one-, two-, and three-loop
levels. It is clear from this figure that NARC possesses the higher
loop stability. The brief description of the properties of the NARC
is given in the next section.

\section{Properties of the New Analytic Running Coupling}

     One of most important features of the new analytic running
coupling is that it incorporates both the asymptotic freedom
behavior and the IR enhancement in a single expression. It was
demonstrated \cite{AlekArbu} that such a behavior of the invariant
charge is in agreement with the Schwinger--Dyson equation. It is
worth noting here that the additional parameters are not introduced
in the theory.

     The consistent continuation of the new analytic running coupling
to the timelike region has been performed recently.\cite{NARCSTR} By
making use of the function $N(a)$ (see Eq.\ (14) in Ref.~8) the
running coupling (\ref{NARCOneLoop}) was presented in the
renorminvariant form and the relevant $\beta$ function was derived:
\begin{equation}
\label{NABeta}
^{\mbox{\tiny N}}\!\widetilde{\beta}
^{\mbox{\scriptsize (1)}}_{\mbox{\scriptsize an}}(a) =
\frac{d\,\ln a(\mu^2)}{d\,\ln\mu^2} =
\frac{1 - N(a)/a}{\ln \left[N(a)\right]}, \quad
a(\mu^2)\equiv\myaanit(\mu^2).
\end{equation}
\begin{figure}[ht]
\vspace*{-5mm}
\centerline{\epsfig{file=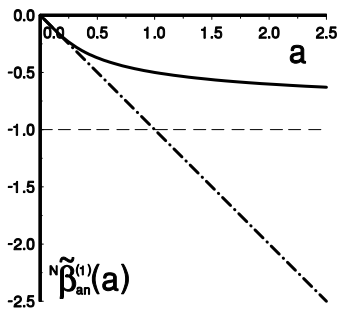, width=45mm}}
\caption{The $\beta$ function corresponding to the one-loop new
analytic running coupling (solid curve). The perturbative result
is shown as the dot-dashed line.}
\label{Fig:NABeta1L}
\end{figure}

     Figure \ref{Fig:NABeta1L} shows the $\beta$ function
(\ref{NABeta}) and the perturbative result corresponding to the
one-loop perturbative running coupling $\widetilde{\alpha}^{(1)}
_{\mbox{\scriptsize s}}(q^2)=1/\ln z$.  The $\beta$ function
(\ref{NABeta}) coincides with the perturbative result
$\widetilde{\beta}^{(1)}_{\mbox{\scriptsize s}}(a)=-a$ in the region
of small values of invariant charge and tends to $-1$ at large $a$.
This results in the IR enhancement of the running coupling, namely,
$\alpha(q^2) \sim 1/q^2$ when $q^2 \to 0$. At the higher loop
levels the $\beta$ function has the same asymptotics. Thus the NARC
possesses the universal IR and UV behavior at any loop level. The
$\beta$ function (\ref{NABeta}) in the region of small $a$ acquires
the form $^{\mbox{\tiny N}}\!\widetilde{\beta} ^{\mbox{\scriptsize
(1)}}_{\mbox{\scriptsize an}}(a) = \widetilde{\beta}^{(1)}
_{\mbox{\scriptsize s}}(a) + O\left[a^{-1} \exp(-a^{-1})\right]$,
which reveals its intrinsically nonperturbative nature. The detailed
description of the properties of the NARC can be found in Ref.~8.

\section{Applications of the New Analytic Running Coupling}

     For verification of the consistency of the model proposed one
should turn to the applications. Since we are working in the
framework of a nonperturbative approach, the study of nonperturbative
phenomena is of a primary interest. The most exciting problem here
is the quark confinement.

     It has been shown \cite{PRD,Vienna} that the new analytic
running coupling (\ref{NARCOneLoop}) explicitly leads to the rising
at large distances static quark-antiquark potential {\it without
invoking any additional assumptions}:
\begin{equation}
^NV(r) \sim \frac{8\pi}{3\beta_0}\Lambda\,\frac{1}{2}
\frac{\Lambda r}{\ln(\Lambda r)}, \quad r \to \infty.
\end{equation}
For the practical purposes it is worth using the approximating
function $U(r)$ for the interquark potential (see Eq.\ (31) in
Ref.~5).  The comparison of $U(r)$ with the Cornell
phenomenological potential $^CV(r)=-4a\,/\,(3r)+\sigma r,\,$
$a=0.39,\,$ $\sigma=0.182\,$GeV$^2$, as well as with the lattice data
\cite{Bali} shows their almost complete coincidence.
\bigskip
\noindent
\begin{center}
\begin{tabular}{cc}
\parbox[t]{55mm}{
\centerline{\epsfig{file=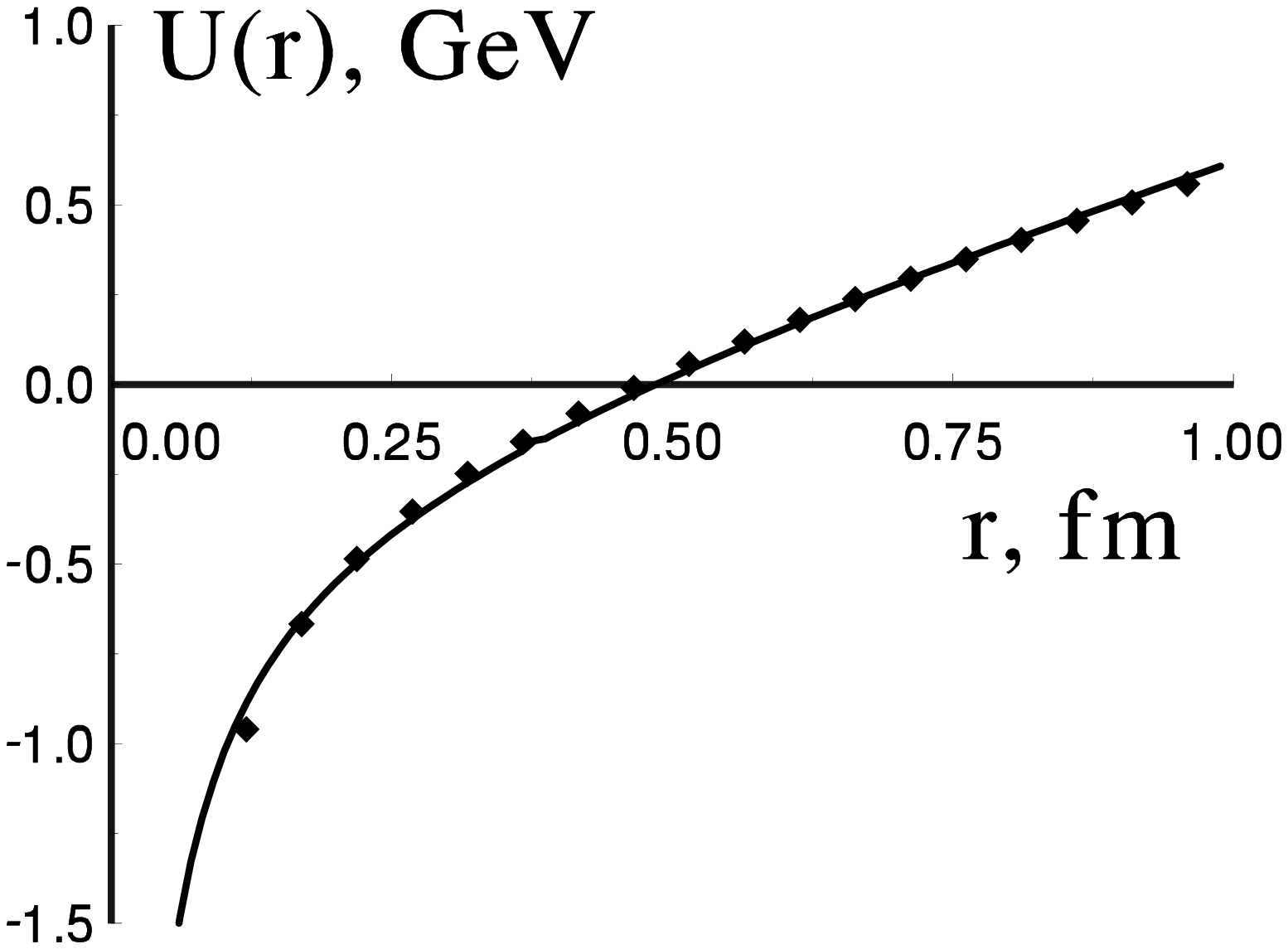, width=52.5mm}}
\parbox[t]{52.5mm}{\footnotesize Figure 3: Comparison of the
potential $U(r)$ (solid curve) with the Cornell potential
($\protect\diamond$); $\Lambda=600\,$MeV, $n_{\mbox{\tiny f}}=3$.}
}
&
\parbox[t]{55mm}{
\centerline{\epsfig{file=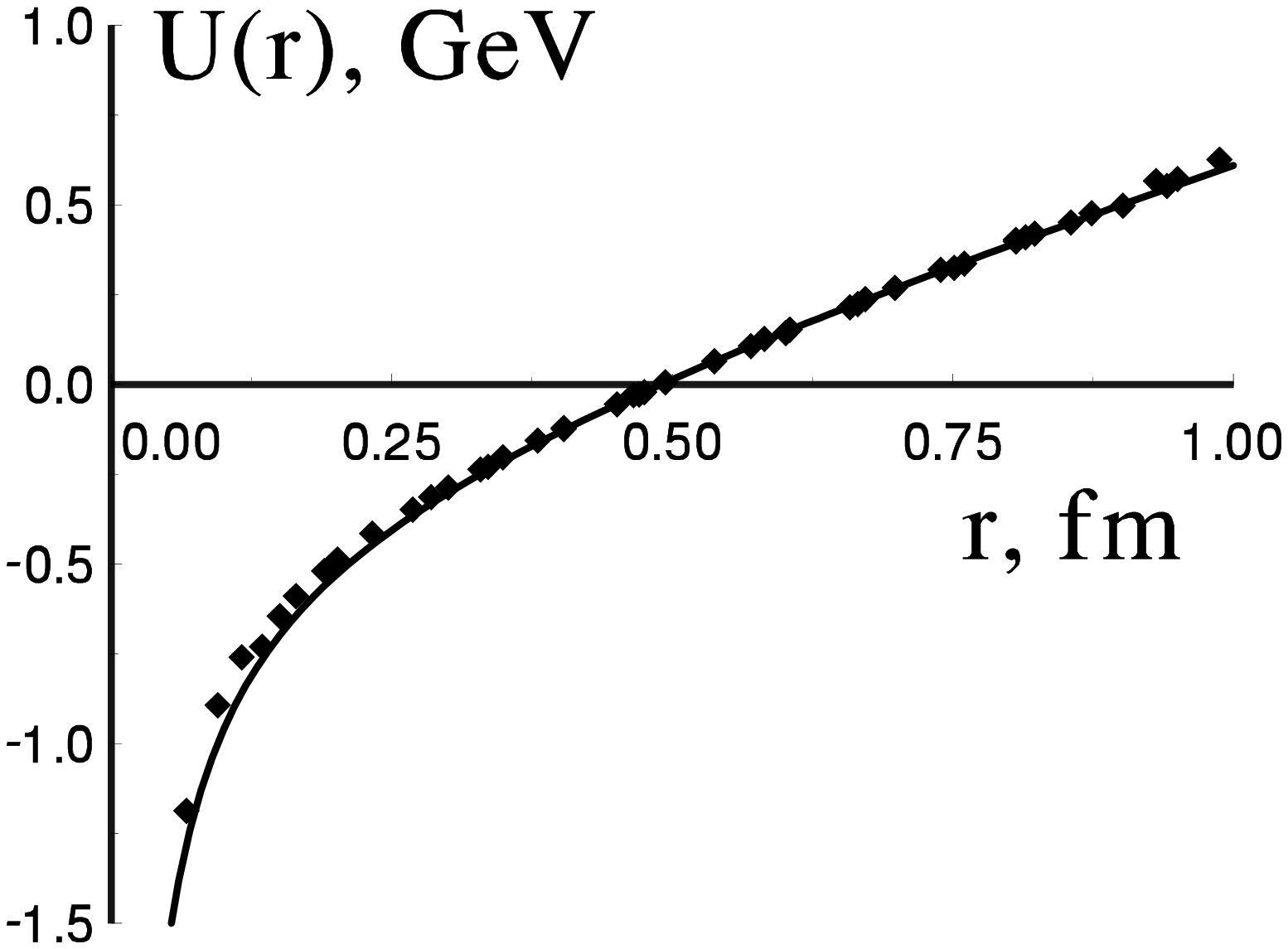, width=52.5mm}}
\parbox[t]{52.5mm}{\footnotesize Figure 4: Comparison of the
potential $U(r)$ (solid curve) with the lattice data
($\protect\diamond$); $\Lambda=580\,$MeV, $n_{\mbox{\tiny f}}=3$.}
}
\end{tabular}
\end{center}
\vspace{2.5mm}

\noindent
It should be mentioned here that the normalization \cite{Bali}
$U(r_0)=0,\,$ $r_0=0.49\,$fm was used, so that \LQCD was the only
varied parameter here.

     By making use of the one-loop new analytic running coupling the
estimation of the parameter \LQCD has been performed recently by
calculation of the value of gluon condensate $K \simeq (0.36\,
\mbox{GeV})^4$. This gave $\Lambda = (650 \pm 50)\,$MeV, which is
close to the previous estimation.

     It turns out that the new analytic running coupling being
applied to the standard perturbative phenomena provides the similar
values of the parameter~$\Lambda$. So, the tentative estimations are
the following.  The $e^{+}e^{-} \to $ hadrons annihilation gives
$\Lambda = 490 \pm 80\,$MeV, and the $\tau$ lepton decay gives
$\Lambda = 520 \pm 65\,$MeV (light quark mass $m \simeq [2.5 \div
6.0]\,$MeV is used here). These values of $\Lambda$ correspond to the
one-loop level with three active quarks.

     Thus, there is a consistent estimation of the parameter \LQCD in
the framework of the approach developed: $\Lambda \simeq 550\,$MeV
(one-loop level, $n_{\mbox{\scriptsize f}}=3$). This testifies that
the new analytic running coupling combines in a consistent way both
nonperturbative and perturbative behavior of QCD.

\section{Conclusions}

     A new model for the QCD analytic running coupling has been
proposed.\cite{PRD} It was presented explicitly in the
renorminvariant form and the relevant $\beta$ function was
derived.\cite{MPLA} The consistent continuation of the NARC
to the timelike region is performed.\cite{NARCSTR} The NARC possesses
a number of appealing features. Namely, there are no unphysical
singularities at any loop level. The IR enhancement and the UV
asymptotic freedom are incorporated in a single expression.  At any
loop level the universal IR behavior ($\alpha \sim 1/q^2$) is
reproduced. The additional parameters are not introduced in the
theory. This approach possesses a good loop and scheme stability. The
confining static quark-antiquark potential is derived {\it without
invoking any additional assumptions}.\cite{PRD,Vienna} There is a
consistent estimation of the parameter \LQCD in the framework of the
current approach.  All this implies that the new analytic running
coupling substantially incorporates in a consistent way perturbative
and nonperturbative behavior of Quantum Chromodynamics.

\section*{Acknowledgments}

     The partial support of RFBR (Grant No.\ 00-15-96691) is
appreciated.

\end{document}